\documentclass[twocolumn,showpacs, pre, aps,secnumarabic,nobalancelastpage,amsmath,amssymb,nofootinbib]{revtex4}

\usepackage{graphics}
\usepackage{graphicx}
\usepackage{amsfonts}
\usepackage{textcomp}
\usepackage{amssymb}
\usepackage{amsmath}
\usepackage{color}
\usepackage{ulem}

\begin{document}

\title{Heterogeneous Force Chains in Cellularized Biopolymer Networks}

\author{Long Liang$^1$, Christopher Jones$^2$, Bo Sun$^2$, and Yang Jiao$^{1,3,*}$}

\affiliation{$^1$Department of Physics, Arizona State University,
Tempe, AZ, 85287} \affiliation{$^2$Department of Physics, Oregon
State University, Corvallis, OR, 97331} \affiliation{$^3$Materials
Science and Engineering, Arizona State University, Tempe, AZ
85287}

\email{yang.jiao.2@asu.edu}

\begin{abstract}
Biopolymer Networks play an important role in coordinating and
regulating collective cellular dynamics via a number of signaling
pathways. Here, we investigate the mechanical response of a model
biopolymer network due to the active contraction of embedded
cells. Specifically, a graph (bond-node) model derived from
confocal microscopy data is used to represent the network
microstructure, and cell contraction is modeled by applying
correlated displacements at specific nodes, representing the focal
adhesion sites. A force-based stochastic relaxation method is
employed to obtain force-balanced network under cell contraction.
We find that the majority of the forces are carried by a small
number of heterogeneous force chains emitted from the contracting
cells. The force chains consist of fiber segments that either
possess a high degree of alignment before cell contraction or are
aligned due to the reorientation induced by cell contraction.
Large fluctuations of the forces along different force chains are
observed. Importantly, the decay of the forces along the force
chains is significantly slower than the decay of radially averaged
forces in the system. These results suggest that the fibreous
nature of biopolymer network structure can support long-range
force transmission and thus, long-range mechanical signaling
between cells.

\end{abstract}



\maketitle

\section{Introduction}



The extracellular matrix (ECM) is an interconnected network of biopolymers that 
provides structural support for cells and allows the diffusion of biochemicals within 
tissues. The most abundant component of ECM is type I collagen, a fibrous protein 
responsible for giving the ECM its material stiffness \cite{Darnell2000}. Cells attach 
and move through the ECM using protein complexes that link the ECM to the 
force-generating cell cytoskeleton \cite{Ridley2003}. However, these cell-ECM 
adhesions also act as sensors, sending information to the cell about the structure 
and mechanical properties of the surrounding matrix \cite{Humphrey2014,Werb2014} 
and helping to regulate cell behavior such as motility, morphology, and 
differentiation \cite{Wang97, Janmey2005, Discher2006}. The stiffness 
and the relative alignment of fibers in the network are particularly important to cell function. 
For example, dense and rigid collagen gel can promote growth and progression of cancer cells
and tumors \cite{Weaver2009, Parsons2010}.  Other important examples are durotaxis 
in which cells tend move in the direction of increasing matrix stiffness \cite{Brown2009}, 
and contact guidance in which cells tend to align and move in the direction of fiber 
alignment \cite{Tranquillo1993, Keely2008}.

Cell-ECM interaction is a dynamic process in which the cell actively remodels the 
network \cite{Werb2014, Erler2011} and these effects can propogate over long distances, 
even affecting the bulk propeties of the network \cite{reorientation14, Howard2002}. Specifically, 
tension exerted by the cells can align the fibers in the network leading 
to long range force transmission \cite{reorientation14, ma2013, Liphardt2014}.
Fiber mediated stresses can trigger mechano-sensitive pathways of distant cells affecting 
behaviors such as force generation \cite{Munro2011,Zallen2009} and cell-ECM 
adhesions \cite{Matsumura2004} and leading to diverse collective behaviors \cite{Forgacs2003}.
This coupling of cells provides a means
for mechanical communication and plays an 
important role in regulating and coordinating collective
cellular dynamics in a wide range of biophysical processes, such as
morphogenesis, tissue regeneration, and immune
response, as well as diseases such as muscular dystrophy, fibrosis, and cancer \cite{Werb2014,
Gilmour2009, Chen2009, Erler2011, Werb1998, Lammerding2009}. 




Due to their effect on cell behavior and communication, 
a significant amount of work has been carried out to
characterize the structural and physical properties of biopolymer
networks. Traditional morphological descriptors for such networks
include the distribution of fiber length \cite{msu}, porosity
\cite{Prendergast2007, Kaufman2010}, pore-size distribution
\cite{ogston1958, taka03, mick2008, jiao2012} and turbidity
\cite{Harbin2000, Kaufman2014}, which are mainly bulk averaged
properties. Recently, local topological and geometrical statistics
such as distribution of number of fibers at a cross-link node
(i.e., valency number) and relative fiber orientations (i.e.,
direction cosine) are employed to successfully reconstruct type I
collagen (COL-I) network computationally \cite{lind2010}. Higher
order spatial fluctuations has also been utilized to characterize
the evolution of COL-I network during gelation process
\cite{jones_spatial-temporal_2014}. In addition, the transport
properties (e.g., macromolecule diffusivity) \cite{jain87,
ogston73, joha91a, joha91b, joha91c, joha93b, ledd06, erik08,
styl10, diff1, jiao2012} and mechanical properties (e.g., elastic
moduli, bulk rheology, stress distribution, etc.) \cite{mech2,
bulk1, lind2010, mech1, ma2013, mech3, Munro2011, hanqing,
zhigang, natphys, reorientation14} of biopolymer network, which
are respectively crucial to the chemical and mechanical signalling
between the cells, strongly depend on the network microstructure.

In most studies, the biopolymer network is treated as a material
system with no biological cells embedded. However, it is known
that cellular interaction with COL-I can induce dynamic remodeling
of the network. These cellular scale dynamics propagate to much
longer range and affect the bulk properties of the network
 \cite{Howard2002}, as well as the behavior of other cells through
mechano-sensitive pathways and machineries \cite{Erler2011,
Werb1998, Lammerding2009}. In addition, fiber mediated
stress-regulated force generation \cite{Zallen2009, Munro2011} and
stress-reinforced cell-ECM adhesion \cite{Matsumura2004} provide
internal feedback mechanisms that support the emergence of diverse
collective behaviors \cite{Forgacs2003}. As a result,
understanding the homeostasis of cellularized biopolymer network
is an essential step towards the understanding of complex,
self-organized multicellular dynamics \cite{Chen2009}.

Recently, continuum models of cellularized collagen network have
been developed to investigate the fiber-mediated mechanical
coupling between the cells \cite{ma2013, Sahai2013}. Specifically,
high-resolution 2D confocal images of cellularized ECM are
thresholded to generate a three-phase heterogeneous system
including a cell phase, a fiber phase and an interstitial fluid
phase \cite{ma2013}. Finite element analysis is then employed to
obtain stress distribution in the ECM. It has been shown that
explicitly considering the fibreous nature of ECM is necessary to
capture the force transmission between cell pairs \cite{ma2013}
and the invasion of cancer cells in ECM with well defined geometry
 \cite{Sahai2013}. Such behaviors can not be correctly reproduced
using either linear or nonlinear homogeneous ECM models. Despite
all these insights, the continuum models do not explicitly
incorporate cell induced fiber re-orientation and other structural
remodeling mechanisms for cellularized ECM, and thus might not
accurately reveal the pathways for force transmission in the
system \cite{reorientation14}.


In this paper, we systematically investigated the mechanical
behavior of cellularized ECM, in particular, fiber-mediated
transmission of forces generated by active contraction of embedded
cells. The biopolymer network is represent by a graph (i.e.,
bond-node) model derived from confocal microscopy data
 \cite{lind2010}. The cell contraction is modeled by applying
correlated displacements at specific nodes (representing the focal
adhesion sites), e.g., displacing the nodes towards a common
center. A force-based stochastic relaxation method is employed to
obtain force-balanced network under cell contraction. We find that
the majority of the forces are carried by a small number of
heterogeneous force chains emitted from the contracting cells. The
force chains consist of fiber segments that either possess a high
degree of alignment before cell contraction or those that are
aligned due to the reorientation induced by the cell contraction.
Fluctuations of the forces along different force chains are
observed. Importantly, the decay of the forces along the force
chains are significantly slower than the decay of radially
averaged forces. These results suggest that network structure
could support long-range force transmission and thus, long-range
mechanical signalling between cells.

The rest of the paper is organized as follows: In Sec.II, we
describe our model of biopolymer network and embedded cells. In
Sec.III, we provide and validate the force-based relaxation
method that minimizes the strain energy of the network to obtain
force-balanced network configuration. In Sec.IV, we discuss the
properties of stressed biopolymer network due to the contraction
of a single cell. In Sec.V, we present results of the mechanical
response of biopolymer network due to the simultaneous contraction
of a pair of cells. In Sec.VI, we provide concluding remarks.

\section{Network and Cell Models}



\begin{figure} [htp]
\begin{center}
$\begin{array}{c@{\hspace{0.5cm}}c}\\
\includegraphics[height=3.25cm,keepaspectratio]{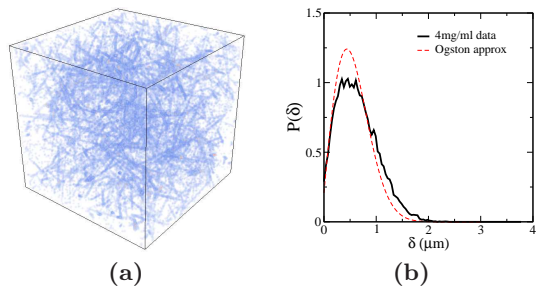} &
\includegraphics[height=3.25cm,keepaspectratio]{fig1b.eps} \\
\mbox{\bf (a)} & \mbox{\bf (b)}
\end{array}$
\end{center}
\label{fig1_pore_size} \caption{(Color online). Microstructure of
the model biopolymer network studied in this paper (a) and the
associated pore-size distribution function $P(\delta)$ (b). The
linear size of simulation box shown in (a) is 30 $\mu$m.}
\end{figure}


The biopolymer network of interest is represented as a graph,
i.e., a collection of bonds and nodes \cite{lind2010}. In this
model, each bond represents a fiber and each node represents a
cross link, see Fig. 1(a) \ref{fig1_pore_size}. The graph
representation is obtained by thresholding and skeletonizing a 3D
stack of confocal microscopy images of type I collagen at density
of 4.0 mg/ml. The average fiber length $\bar{d^0} = 1.28 \mu$m. A
cubic simulation box with a linear size of $30 \mu$m is used,
which contains 5,000 nodes and 8,000 bonds. The pore-size
distribution function of the network is shown in Fig. 1(b)
\ref{fig1_pore_size} \cite{jiao12}. Also shown is the Ogston
approximation for the pore size distribution, i.e.,
\begin{equation}
\label{eq_POgston} P(\delta) = \frac{2\phi_2
(\delta+a)}{a^2}e^{-\phi_2 (\delta+a)^2/a^2},
\end{equation}
where $\phi_2 = 1-\phi_1$ is the volume fraction of the fibers and
$a$ is the fiber radius. We note that (\ref{eq_POgston}) is
derived for networks with very long and stiff fibers, and thus,
overestimates the occurrence probability of intermediate sized
pores \cite{jiao12}.


In order to study the mechanical behavior of the network, we need
to specify the properties of individual fibers. In this paper, we
assume that the fibers possess short persistent lengths, i.e., the
bending modulus of the fibers is significantly smaller than the
elongation modulus. In addition, we consider that cell contraction
can only generate small forces and thus, the system is in the
linear elastic regime \cite{ma2013} and the effects of
interstitial fluid, which quickly dissipates the kinetic energy
generated due to cell contraction, are not explicitly considered.
The mechanically equilibrated network possesses the minimal
elastic (strain/stress) energy. The elongation modulus of the
fibers is set to be $EA = 8\times 10^{-7} N$  \cite{lind2010},
where $E$ is the Young's modulus of collagen and $A$ is the
cross-section area of the fiber. The elastic energy of a fiber
$\ell_{ij}$ defined by nodes $i$ and $j$) is a quadric function of
fiber elongation, i.e.,
\begin{equation}
\epsilon_{ij} = \frac{EA}{2 d_{ij}^{0}}\cdot (d_{ij}-d_{ij}^{0})^2
= k_{ij}\cdot(d_{ij}-d_{ij}^{0})^2,
\end{equation}
where the spring constant of fiber $k_{ij} = EA/d^0_{ij}$ is
inversely proportional to its original length $d_{ij}^{0}$. For
computational convenience, the coordinates of the nodes inside the
$30\mu m \times 30\mu m \times 30\mu m$ box are rescaled to fit in
a $1\times1\times1$ simulation box. In the subsequent
calculations, the unit of length is thus chosen to be the length
of the simulation box. We note that the results can be easily
rescaled to the actual units in order to compare with experimental
measurements. The total elastic energy associated with the fiber
network is given by
\begin{equation}
\label{E_G} E_{\mbox{\tiny G}} = \sum\limits_{<i\ j>}\epsilon_{ij}
\end{equation}
where the summation is over all pairs of connected nodes (i.e.,
all fibers $\ell_{ij}$).
\newline

\begin{figure}[!h]
\includegraphics[height=5.0cm,keepaspectratio]{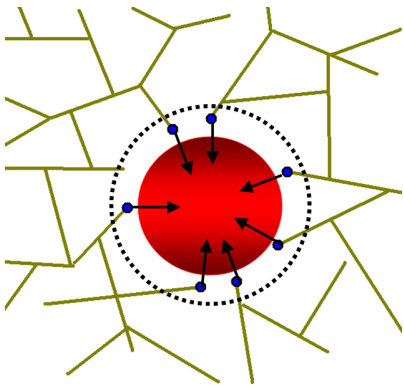}
\caption{(Color online). A schematic illustration of our cell
contract model. We model an embedded cell as a sphere with radius
$R_c$ (red circle) centered at a prescribed location in the
network (dark yellow lines). The nodes of the network within a
certain distance $\delta R$ to the cell surface (i.e., $R_c+\delta
R$ to the cell center) are considered as ``coarse-grained'' focal
adhesion sites (blue dots). Cell contraction is then modeled by
displacing the focal adhesion nodes towards the cell center with a
prescribed magnitude.} \label{fig:cell_contract}
\end{figure}

We model an embedded cell as a spherical region with radius $R_c$
centered at a prescribed location in the network. Although actual
cells generally possess much more complex morphology, such a
simple shape is sufficient for our current study, which focuses on
heterogeneous force chains generated due to cell contraction. The
nodes of the network within a certain distance $\delta R$ to the
cell surface (i.e., $R_c+\delta R$ to the cell center) are
considered as ``coarse-grained'' focal adhesion sites, through
which the cell is mechanically coupled with the network
\cite{plotnikov2013guiding}. Cell contraction is then modeled by
displacing the focal adhesion nodes towards the cell center with a
prescribed magnitude. In particular, with the boundary nodes kept
fixed (a node is a boundary node if its distance to the box
boundary is smaller than 5\% of the box length), the set of
adhesion nodes are displaced towards to cell center and then kept
fixed. This local perturbation leads to a stressed network and the
resulting force-balanced configuration is obtained via a
force-based stochastic relaxation method (described in detail
below): All of the other nodes (free nodes) are allowed to be
displaced until the entire network settles down in a new
configuration with minimal strain/stress energy minima such that
all free nodes are in a force balanced state.




\section{Force-based Relaxation Method}



Here, we employ a stochastic optimization scheme to find the
force-balanced network configuration under the prescribed local
perturbations that mimic cell contraction. The standard stochastic
optimization procedure consists of a random walk in the
configuration space. Specifically, in order to find the minimal
energy states, a randomly selected node of the network is given a
random small displacement (i.e., a trial move), which leads to
stretching/compression of fibers that connected to this node and
thus, causes a change of total elastic energy $E_{\mbox{\tiny G}}$
of the network. If $E_{\mbox{\tiny G}}$ decreases, the
displacement is accepted, otherwise, it is rejected, i.e., the
steepest descent method is utilized.

In this scheme, the global energy needs to be re-computed after
each trial displacement. Directly re-computing $E_{\mbox{\tiny
G}}$ using Eq. (\ref{E_G}) is both time consuming and
computationally inefficient due to the large number of nodes
(5000) and bonds (8000) in our system, as well as the large number
of trial moves (i.e., $\sim 5000$ moves for each node). Therefore,
a local energy update method \cite{jiao_pre, jiao_pre2, jiao_pnas}
is employed: before displacing the randomly chosen node, the
energy associated with this node, i.e., the ``local energy''
$E_{\mbox{\tiny L}}$, which is defined as the elastic energy of
all the bonds (fibers) connected to this specific node, is
calculated. Because the displacement of a node only affect the
bonds connected to this node, the change of the total energy
$\delta E_{\mbox{\tiny G}}$ is exactly equal to the change this
local energy $\delta E_{\mbox{\tiny L}}$, i.e., $\delta
E_{\mbox{\tiny G}} = \delta E_{\mbox{\tiny L}}$. The acceptance
probability of the trial move $p_{acc}$ is simply given by
\begin{equation}
\label{p_acc} p_{acc} = \left\{{\begin{array}{*{2}c}1, \quad
\delta E_{\mbox{\tiny L}} \le 0 \\\\ 0, \quad\quad \delta
E_{\mbox{\tiny L}}>0.
\end{array}}\right.
\end{equation}
This also enables of the efficient computation of both the global
energy and thus, significantly reduces the computational cost,
i.e., from 8 CPU hrs to 0.6 CPU hrs on our Dell Precision
workstation [with Intel Xeon(R) E5-1603 2.80GHz 4-core CPU and 16
GB Memory).

Furthermore, instead of randomly displacing the nodes, force-based
displacements are used. Specifically, a randomly selected node is
displaced along the direction of the net force on this node, with
the magnitude of the displacement proportional to the magnitude of
the force. For a randomly selected node $i$, the net force ${\bf
F}^i$ at this node with the components $F^i_x$, $F^i_y$ and
$F^i_z$ is calculated by summing the $x$, $y$, $z$ components of
all the forces exerted by all the neighbor nodes on this chosen
node $i$. The magnitude of the force between node $i$ and its
neighbor $k$ connected by a fiber is computed as
\begin{equation}
f_{ik} = \frac{EA}{d^0_{ik}}\cdot (d_{ik}-d^0_{ik}),
\end{equation}
where $E$ is the Young's modulus of the fiber, $A$ is the fiber
cross section area, $d_{ik}$ and $d^0_{ik}$ are respectively the
fiber lengths after node displacement and original fiber length.
Since $f_{ik}$ is along the direction defined by nodes $i$ and
$k$, the $x$ component of $f_{ik}$ can be computed by multiplying
$f_{ik}$ by the $x$ component of the normalized direction vector
${\bf T}^{ik}$ pointing from node $i$ to $k$, i.e.,
\begin{equation}
T^{ik}_x = \frac{(x_k-x_i)}{d_{ik}},
\end{equation}
where $x_i$ and $x_k$ are respectively the $x$ coordinates of node
$i$ and $k$. The $x$ component of the net force at node $i$ is
then given by
\begin{equation}
F^i_x = \sum\limits_{<k>} f_{ik}\times T^{ik}_x.
\end{equation}
The components $F^i_y$ and $F^i_z$ can be calculated in a similar
way, which we do not describe in detail here.

Finally, the node $i$ is displaced along the direction of ${\bf
F}^i$ with a magnitude
\begin{equation}
\delta_i = C\cdot\frac{d^0_{ik}}{EA}\cdot{\bf F}^i,
\end{equation}
where $C$ is random multiplier between [0, 625,000]. The upper
bound is chosen such that the maximal individual displacement is
$\sim 1/500$ average fiber length, which leads to fast convergence
of the optimization. About 5000 trial moves are performed on each
node, with an average success rate of 95\%. The simulation is
terminated when the total energy converge to a stable plateau.

\begin{table*}[t]
\begin{tabular}
{c@{\hspace{0.1cm}}c@{\hspace{0.25cm}}c@{\hspace{0.25cm}}c
@{\hspace{0.25cm}}c} \hline\hline
& force-based local &¡¡random local & force-based global & random
global \\
\hline
$t_c$ & {38 min} & {40 min} & {8hr 41min } & {8hr 59min} \\
\hline
$E^0_{\mbox{\tiny G}}$ & 28.46740 & 28.46740 & 28.46740 & 28.46740 \\
\hline
$E^f_{\mbox{\tiny G}}$ & 6.7716672 & 6.6983984 & 6.7727792 & 6.6982896 \\
\hline
$N_t$ & 25,000,000 & 25,000,000 & 25,000,000 & 25,000,000\\
\hline
$N_s$ & 23,816,643 & 1,143,437 & 23,785,917 & 1,143,636 \\
\hline
$\gamma$ & 95.3\% & 4.57\% & 95.1\% & 4.57\% \\
\hline\hline
\end{tabular}
\caption{Comparison of different stochastic energy minimization
schemes to obtained force-balanced biopolymer network
configurations. $t_c$ is the total CPU computational time
consumed, $E^0_{\mbox{\tiny G}}$ is the initial energy (in $pJ$),
$E^f_{\mbox{\tiny G}}$ is the final energy (in $pJ$), $N_t$ is the
total number of MC moves, $N_s$ is the number of successfully MC
moves, and $\gamma = N_s/N_t$ is the success rate.}
\end{table*}

\begin{figure} [htp]
\begin{center}
$\begin{array}{c@{\hspace{0.25cm}}c}\\
\includegraphics[height=3.0cm,keepaspectratio]{Comparison_of_Energy.eps} &
\includegraphics[height=3.0cm,keepaspectratio]{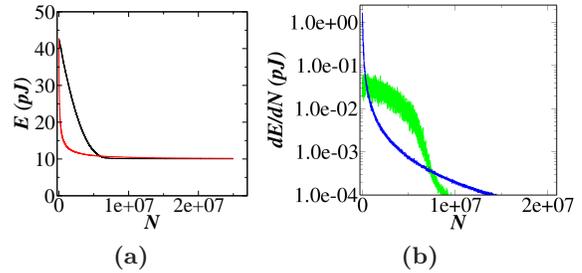} \\
\mbox{\bf (a)} & \mbox{\bf (b)}
\end{array}$
\end{center}
\label{fig_energy_scheme} \caption{(Color online). Comparison of
different optimization procedures. (a) The total energy (in pJ) as
a function of trial move number. The red (or dark gray in print
version) curve corresponds to the force-based method and the black
curve corresponds to the random displacement method. (b) The rate
of change of energy. The blue (or dark gray in print version)
curve corresponds to the force-based method and the green (or light
gray in print version) curve corresponds to the random
displacement method. It can be seen that the force-based scheme
significantly improve convergency of the optimization.}
\end{figure}


Table I compares the efficiency of different optimization schemes
described above. Specifically, the different procedures yield
essentially identical final energy, indicating all of them are
robust in finding the desired (local) energy minimum. However, the
local energy-update scheme accelerates the simulation by 12 times,
and the force-based method possesses a success rate that is 20
times higher than the random displacement approach. The
differences in the force-based approach and random-displacement
approach is also shown in Fig. \ref{fig_energy_scheme}. It can be
clearly seen that the force-based scheme significantly improve
convergency of the optimization.

\section{Single-cell Contraction Results}

In this section, the force-based relaxation method is applied to
study the mechanical response of a biopolymer network due to the
contraction of a virtual cell embedded in the network. In
particular, a spherical cell is placed in the center of the
simulation box, i.e. at $(15, 15, 15)(\mu m)$, with a radius $R_c
= 4.5 \mu m$ (i.e., a linear size of $9 \mu m$). The contraction
ratio $\Gamma_c$ (defined as the ratio of the cell radius after
and before the contraction) is chosen to be 0.3. We consider the
``cell'' is initially residing inside a stress-free network.
Therefore it is easy for the cell to perturb the network via
contraction, leading to relatively large $\Gamma_c$ used here. The
force-based relaxation method is then employed to find
force-balanced network configuration.

\subsection{Heterogeneous distribution of forces on fibers}

\begin{figure}[!h]
$\begin{array}{c}
\includegraphics[height=5.0cm,keepaspectratio]{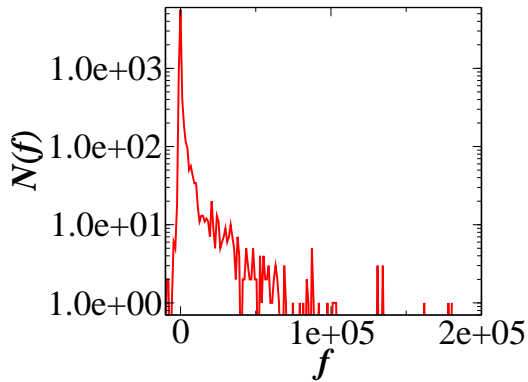}\\
\end{array}$
\caption{(Color online). Distribution of forces (in $pN$) on the
fibers. The majority of fibers ($\sim 5000$ out of 8~000 fiber in
our system) carries very small forces. There are only a small
number of fibers that carry the majority of the forces $\sim 10^4$
$pN$. Such fibers form linear structures, which we refer to as
force chains (see Fig. \ref{fig:force_chain_of_single_cell}).}
\label{fig:force_distribution}
\end{figure}

The distribution of forces on the fibers is shown in Fig.
\ref{fig:force_distribution}. It can be clearly seen that the
majority of fibers ($\sim 5~000$ out of 8~000 fibers in our
system) carries very small forces (i.e., virtually zero). There
are only a small number of fibers that carry the majority of the
forces $\sim 10^4$ $pN$, which leads to a highly heterogeneous
distribution of forces in the network.


\subsection{Identifying force chains}

\begin{figure}[!h]
$\begin{array}{c}
\includegraphics[height=5.0cm,keepaspectratio]{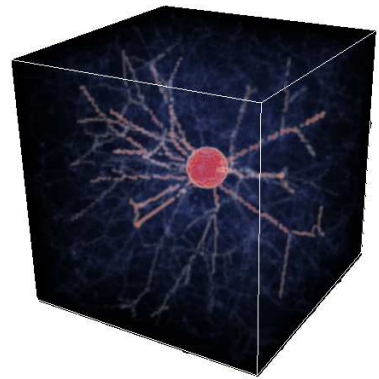}\\
{\mbox{\bf (a)}}\\\\
\includegraphics[height=4.5cm,keepaspectratio]{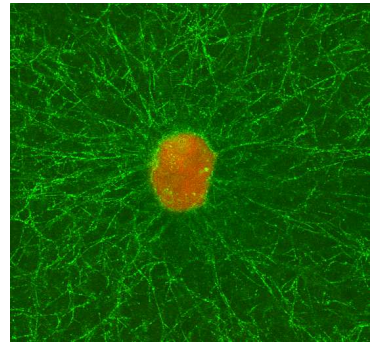}\\
{\mbox{\bf (b)}}\\\\
\end{array}$
\caption{(Color online). Force chains in the biopolymer network
due to the contraction of an embedded spherical cell. (a)
Simulation result: the fibers carrying large forces are
highlighted with red color. It can be clearly seen that there are
chain-like structures emitted from the contracted cell, consisting
of fibers bearing very large forces ($\sim 10^4$ $pN$). (b)
Experimental validation: A contracting NIH 3T3 cell with an almost
spherical morphology in collagen I gel with a concentration of
1.5mg/ml. Similar linear chain-like structures of collagen fibers
emitting from the cell can be clearly observed.}
\label{fig:force_chain_of_single_cell}
\end{figure}

To understand how the local perturbation due to cell contraction
is propagated throughout the system, we investigate the spatial
correlations among large-force-bearing fibers and identify
well-defined structures formed by these fibers. Specifically, we
visualize the entire system in a $100\times100\times100$
voxel-based box, with different color code corresponding to
different stress states of the fibers. The gradual change of color
from blue to red indicates the increase of the forces carried by
the fibers. Surprisingly, we find that there are well-defined
chain-like structures composed of large-force-bearing fibers
(i.e., red fibers) in the network, which are emitting from the
contracted cell, i.e., the spherical region centered at $(15, 15,
15)$ ($\mu m$) and protruding outwards along the radial direction,
see Fig. \ref{fig:force_chain_of_single_cell}(a). We refer to
these chain-like structures as ``force chains'' in the biopolymer
network. Figure \ref{fig:force_chain_of_single_cell}(b) shows the
snapshot of a contracting NIH 3T3 cell with an almost spherical
morphology in collagen I gel with a concentration of 1.5mg/ml.
Similar linear chain-like structures of collagen fibers emitting
from the cell can be clearly observed, which validates our
simulation.

Starting from the contracted spherical cell, we track the
propagation of forces (generated by the cell) along the force
chains. In particular, the origin of a force chain is associated
with a fiber that is directly connected to the cell, i.e., with
one node as the coarsen-grained ``focal adhesion'' site, points
outwardly from the cell and carries a force larger than a
threshold value (e.g., $10^4$ pN). The next segment of the force
chain is then identify as the fiber that shares a common node of
the origin fiber, points outwardly from the cell and carries a
force that is larger than all of the other fibers connected to the
shared node. This process is repeated until the force chain
extends to the boundary of the box. Although it is not the only
way to identify force chains, our procedure identifies force
chains with largest linear extent by imposing the condition that
the segments of the force chain always point outwardly from the
cell. This allows us to investigate the range of fiber-mediated
mechanical signaling between cells.



\begin{figure}[!h]
\includegraphics[height=5.0cm,keepaspectratio]{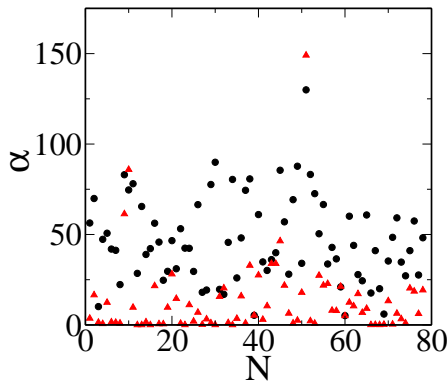}
\caption{(Color online). Comparison of the angles $\alpha$ between
two successive fiber segments along the force chains before (black
dots) and after (red dots) the cell contraction. This comparison
reveals two mechanisms for force-chain formation, i.e., fiber
re-orientation and selection of pre-existing chain-like structures
in the network, as discussed in the text.} \label{fig:realignment}
\end{figure}



How does such chain-like structures arise in response to the
perturbation due to cell contraction? To investigate the mechanism
for the formation of such linear structure, we compare the angles
$\alpha$ between two successive fiber segments along the force
chains, before and after the cell contraction. Figure
\ref{fig:realignment} shows the comparison. It can be clearly seen
that most of the angles $\alpha$ after cell contraction (shown as
red dots) are relatively small (i.e., less than 20 degrees), which
is consistent with observed linear chain-like structures. However,
a significant fraction of $\alpha$ ($\sim 70\%$) before
contraction (shown as black dots) possesses large values (e.g.,
$>40$ degrees). This suggests that roughly $70\%$ fibers undergo
large re-orientation due to cell contraction in order to support
the propagation of forces through the linear force chains. We note
this result is consistent with those reported in Ref.
\cite{reorientation14}. The remaining angles, whose values before
the contraction are relatively small, do not significant change
after the perturbation. This indicates that certain pre-existing
chain-like structures in the non-stressed network can also be
selected to carry large forces and contribute to the force chains.
The relative contributions of the aforementioned two mechanisms
for force-chain formation, i.e., fiber orientation and selection
of pre-existing linear structures, generally depends on the
network of microstructure.

\subsection{Force decay along force chains}

\begin{figure}[!h]
\includegraphics[height=5.0cm,keepaspectratio]{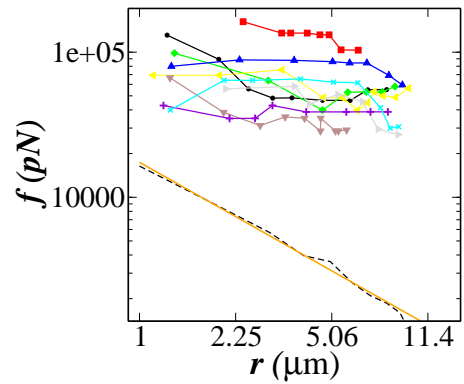}
\caption{(Color online). Comparison of the decay of forces
$f_{C}$(in $pN$) from the contracting cell along the force chains
(solid symbols) and the decay of radially averaged force
$\bar{f}$(in $pN$) in the biopolymer network (dashed line). It can
be clearly seen that the decay of $f_C$ is much slower than decay
of $\bar{f}$. This suggests a possible mechanism for long-range
force transmission in cellularized biopolymer network.}
\label{fig:Comparison of Decay behavior}
\end{figure}

To quantify how the stress propagates along the force chains, the
fiber segments along each force chain have been identified and the
magnitude of the forces on the fibers along each force chain are
obtained. Figure \ref{fig:Comparison of Decay behavior} shows the
comparison of the decay of forces along the force chains
$f_{C}(r)$ and the decay of radially averaged force ${\bar f}(r)$
as the distance $r$ form the contracted cell increases. We note
that ${\bar f}(r)$ is computed by averaging the magnitude of
forces carried by the fiber segments whose centers are in a concentric thin
spherical shell with radius $r$ and thickness $dr$.

It can be clearly seen in Fig. \ref{fig:Comparison of Decay
behavior} that the decay behavior of $f_C(r)$ is distinctly
different than ${\bar f}(r)$. In particular, $f_C(r)$ decays much
slower than ${\bar f}(r)$, although both are characterized by a
power law decay as can be clearly seen in the logarithmic plot.
The orange line shows the linear fit of $\ln{\bar{f}}$ vs.
$\ln{r}$, i.e.,
\begin{equation}
\ln{\bar{f}(r)} = 9.765 - 1.062 \ln{r}.
\end{equation}
The slope 1.062 indicates that the decay of ${\bar{f}(r)}$ can be very well
described via
\begin{equation}
{\bar f}(r) \sim 1/r,
\end{equation}
which is consistent with the prediction from linear elastic theory
\cite{sal_book}. Due the large fluctuations of the forces along
the force chain, the exact decay behavior of $f_C(r)$ is difficult
to extract, but can be approximated via
\begin{equation}
f_C(r) \sim 1/r^\eta,
\end{equation}
where $\eta \in (0.3, 0.5)$. In addition, the magnitude of the
forces along the force chain is much higher than the radially
averaged forces. This is consistent with our observation from the
distribution of forces shown in Fig. \ref{fig:force_distribution},
i.e., the majority of forces are carried by a small number of
fibers constituting the force chains, while the remaining fibers
carry very small forces (see fig5 \& fig7)
\ref{fig:force_chain_of_single_cell}).


\section{Double-cell Contraction Results}


To further understand fiber-mediated mechanical coupling between
cells, we investigate the force chains formed due to the
contraction of two cells that are close to one another. In
particular, two cells with $R_c = 4.5 \mu m$ are placed on the
diagonal line of the cubic simulation box, respectively at{
{$(10.5, 10.5, 10.5) \mu m$ and $(19.5, 19.5, 19.5) \mu m$}. We
note that these points are the trisection points of the body
diagonal of the ``free'' part of the simulation box. A contraction
ratio of $\Gamma_c = 0.5$, instead of 0.3 for the single cell
case, is employed. This is because when more than one cells are
embedded in the network, the system can be considered in a
prestressed state and thus, is more difficult to mechanically
perturb than the single cell case. After the perturbation (i.e.,
simultaneous contraction of both cells), the system is
equilibrated using the force-based relaxation method.
Specifically, each node is given on average 7,000 trial moves with
an average success rate of $97.35\%$.

\begin{figure}[!h]
\includegraphics[height=5.5cm,keepaspectratio]{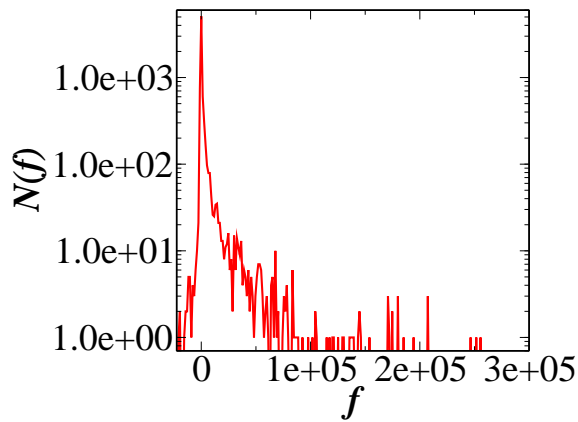}
\caption{(Color online). Distribution of the forces (in $pN$)
carried by the fibers in the biopolymer network with two
contracting cells. It can be seen that the majority of fibers only
carry very small forces close to zero. The larger forces are
mainly carried by fibers forming force chaining connecting the two
contracting cells and emitting outward to the boundary.}
\label{fig:force_distribution2}
\end{figure}

\begin{figure}[!h]
$\begin{array}{c}
\includegraphics[height=5.0cm,keepaspectratio]{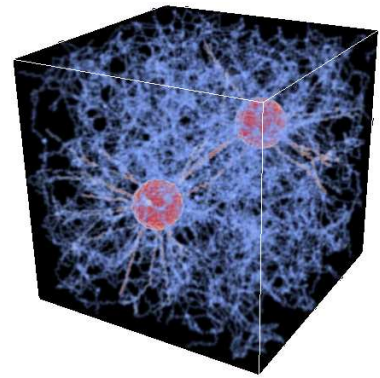}\\
\mbox{\bf {(a)}}\\\\
\includegraphics[height=4.5cm,keepaspectratio]{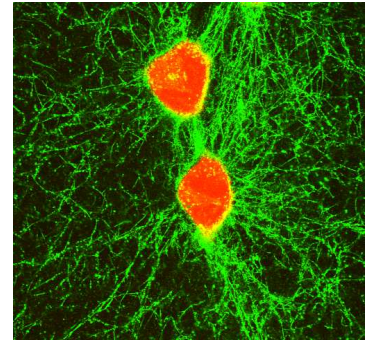}\\
\mbox{\bf {(b)}}\\\\
\end{array}$
 \caption{(Color online). Force chains in the
biopolymer network due to the contraction of two embedded cells.
(a) Simulation result: The fibers carrying large forces are
highlighted with red color. It can be clearly seen that there are
chain-like structures connecting the two cells, consisting of
fibers bearing very large forces ($> 10^5$ $pN$). (b) Experimental
validation: Two contracting NIH 3T3 cells close to one another in
collagen I gel with a concentration of 1.5mg/ml. Similar linear
chain-like structures of collagen fibers connecting the two cells
can be clearly observed.} \label{fig:2 cell case}
\end{figure}

Figure \ref{fig:force_distribution2} shows the distribution of the
forces carried by the fibers. As in the single-cell case, the
majority of fibers only carry very small forces close to zero. The
majority of forces are carried by a small number of fibers which,
as shown below, form linear chain-like structures. Figure
\ref{fig:2 cell case} shows the stressed network due to
contraction of the cells. The color code is same as in the
single-cell case, i.e., the fibers carrying large forces are
highlighted with red color. Two force-chains connecting the two
contracting cells can be clearly identified, both consisting fiber
elements carrying a forces large than $10^6$ pN (see Appendix A).
The fibers in the force chains are aligned very well with the body
diagonal line, on which the two cells are placed. Figure
\ref{fig:2 cell case}(b) shows the snapshot of two contracting NIH
3T3 cells close to one another in collagen I gel with a
concentration of 1.5mg/ml. Similar linear chain-like structures of
collagen fibers connecting the two cells can be clearly observed,
which validates our simulation.

\begin{figure}[!h]
\includegraphics[height=5.0cm,keepaspectratio]{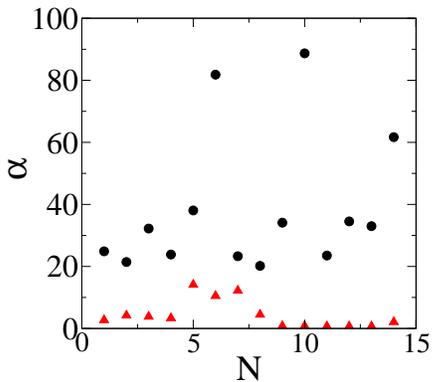}
\caption{(Color online). Comparison the angles between individual
fiber segment along the force-chains and the body diagonal, before
(black dots) and after (red dots) cell contraction. The fiber
segments along the force chains undergo a significant
reorientation to align with the body diagonal, as discussed in the
text. We note that the angles $\alpha$ shown here are with respect
to the body diagonal of the simulation box, rather than the angles
between successive fiber segments in the single-cell case.}
\label{fig:realignment2}
\end{figure}

To understand how such linear-structures form, we compute and
compare the angles between individual fiber segments along the
force-chains and the body diagonal, before and after cell
contraction. Figure \ref{fig:realignment2} shows the comparison.
It can be clearly seen that the fiber segments along the two most
prominent force chains between the two cells undergo a significant
reorientation to align with the body diagonal. We note that in
this configuration, the largest forces occur along the body
diagonal connecting the 2 cells, which also drive the
re-orientation of the fibers. In addition, the fiber segments in
the force chain carrying large forces undergo a more significant
re-orientation (see Appendix A), to get better aligned with the
body diagonal in order to resist the larger forces. The
``selection'' mechanism for force-chain formation is not dominant
in the two-cell case. This is because in our current network, the
probability of finding a chain of fibers all perfectly aligned
with the body diagonal is very small. Finally, we note that since
the two-cell configuration does not possess spherical symmetry, we
therefore do not characterize the radial decay of the forces,
which is not well defined. The magnitude of forces carried by the
fiber segments along the force chains are given in Appendix A.

\section{Conclusions and Discussion}


We have numerically investigated the mechanical response of a
model biopolymer network due to the active contraction of embedded
cells. Cell contractions are modeled by applying correlated
displacements at specific nodes, representing the focal adhesion
sites. A force-based stochastic relaxation method is employed to
obtain force-balanced network under cell contraction. We find that
the majority of the forces are carried by a small number of
heterogeneous force chains emitted from the contracting cells. The
force chains consist of fiber segments that either possess a high
degree of alignment before cell contraction or are aligned due to
the reorientation induced by cell contraction. Large fluctuations
of the forces along different force chains are observed [c.f. Fig.
7]. In addition, the decay of the forces along the force chains
are significantly slower than the decay of concentric averaged
forces in the system. These results suggest that the fibreous
nature of biopolymer network structure can support long-range
force transmission and thus, long-range mechanical signaling
between cells \cite{wang2009mechanotransduction}. One of the adavantages
of Mechano-signal transduction over Chemcial signaling is that
it could be upto 40 times faster than the diffusion-based, as is
reviewed in \cite{wang2008}.

We note that force chains are also apparent in granular materials,
which is non-equilibrium state of matter composed of macroscopic
particles with strong repulsive interactions
\cite{peters2005characterization}. Compared to the typical force
chains in granular materials, the force chains we identified in
the perturbed network have a much better linearity, i.e., they
either aligned radially outwards in the single-cell case or align
the body diagonal in the two-cell case. In addition, the forces
carried by the fiber segments are stretching forces, while those
carried by the individual grains are compressive in nature. The
force chains in these two distinct systems also share some similar
characteristics. For example, in both systems, the force chains
carry the majority of the forces, which are orders of magnitude
higher than the average forces carried by a fiber/grain. This
implies that the formation of force chains could be a universal
mechanism for stress dissipation in disordered systems composed a
large number of individual building blocks with interaction rules in between.

Also notably, from Figure 7, we can also see the ``durality'' of 
the collagen network system: on one hand, if we are interested 
in the bulk property of the system, it's totally fine to treat the 
collagen network as homogeneous in the same way as all the other continuum materials, 
which is evident from the perfect 1/r overall decay of the stress field
, on the other hand, if we are more intrigued by
the underlying events happening at even smaller scales, the 
fiberous nature of collagen system can not neglect.

Finally, we note that our biopolymer network model only considers
fiber elongation and does not explicitly take into account the
effect of bending. This assumption works perfectly fine for fibers
with short persistent lengths, but is not true for fibers with
long persistent lengths. In future work, the effects of fiber
bending will be explicitly investigated. In addition, the effects
of fiber alignment in the original non-stressed network work will
also be studied. It is expected that in a highly-orientated fiber
network, the ``selection mechanism'' will dominant for force-chain
formations, and thus, significantly biases the force propagation
as well as the cell migration inside.





\begin{acknowledgments}
L. L. and Y. J. thank Arizona State University for the generous
start-up funds.
\end{acknowledgments}

\newpage
\begin{appendix}
\section{Forces on force chain elements in double-cell
configuration}
\begin{table}[htbp]
\caption{Realignment of the fibers (i.e., change of the angle
$\alpha$ (degree) between two successive fiber segments before and
after contraction) in the force chain between the 2 cells and the
forces (pN) carried by the fiber segments.}
    \begin{tabular}{|r|r|r|r|}
    \hline
    Element & $\alpha$ Before & $\alpha$ After & Force \\ \hline
    \multicolumn{4}{c}{Force Chain 1} \\ \hline
    0        & 24.87 & 2.71  & 174,334.2 \\ \hline
    1       & 21.43 & 4.21  & 182,937.9 \\ \hline
    2       & 32.22 & 3.83  & 182,290.5 \\ \hline
    3       & 23.85 & 3.34  & 181,654.4 \\ \hline
    4        & 38.05 & 14.12 & 164,516.6 \\ \hline
    5       & 81.81 & 10.51 & 152,255.8 \\ \hline
    6       & 23.30 & 12.23 & 169,661.7 \\ \hline
    \multicolumn{4}{c}{Force Chain 2} \\  \hline
    0       & 20.17 & 4.49  & 315,013.8 \\ \hline
    1        & 34.11 & 0.81  & 219,117.7 \\ \hline
    2       & 88.69 & 0.81  & 219,118.4 \\ \hline
    3        & 23.52 & 0.71  & 224,975.0 \\ \hline
    4        & 34.51 & 0.72  & 224,988.8 \\ \hline
    5       & 32.98 & 0.72  & 224,989.5 \\ \hline
    6      & 61.68 & 2.10  & 308,953.9 \\ \hline
    \end{tabular}%
  \label{tab:addlabel}%
\end{table}%

\end{appendix}



\providecommand{\noopsort}[1]{}\providecommand{\singleletter}[1]{#1}%

\end{document}